\documentclass[dvips, usenatbib]{mn2e}
\usepackage[T1]{fontenc}
\usepackage[latin9]{inputenc}
\usepackage{amsmath}
\usepackage{amsfonts}
\usepackage{amssymb}
\usepackage[dvips]{graphicx}
\usepackage{booktabs}
\usepackage{tabularx}
\usepackage{natbib}
\usepackage{hyperref}            
\newcommand{\OM}{\Omega_{\Lambda, 0}}

\begin{document}

\title[A Two-Parameter Matching Scheme for Galaxies and Haloes]{A Two-Parameter 
Matching Scheme for Massive Galaxies and Dark Matter Haloes}

\author[A. Kulier \& J. P. Ostriker]{
Andrea Kulier$^{1}$\thanks{E-mail: \href{mailto:akulier@princeton.edu}{akulier@princeton.edu}} and Jeremiah P. Ostriker$^{1,2}$\\
$^{1}$Department of Astrophysical Sciences, Princeton University, Princeton, NJ 08544, USA\\
$^{2}$Department of Astronomy, Columbia University, NYC, NY 10027, USA}

\date{\today}

\pagerange{\pageref{firstpage}--\pageref{lastpage}} \pubyear{2015}

\maketitle

\label{firstpage}

\begin{abstract}
Halo Abundance Matching has been used to construct a one-parameter
mapping between galaxies and dark matter haloes by assuming that halo mass and galaxy luminosity
(or stellar mass) are monotonically related. While this approach has been reasonably successful,
it is known that galaxies must be described by at least two parameters, as can be seen
from the two-parameter Fundamental Plane on which massive early-type galaxies lie.
In this paper, we derive a connection between initial dark matter density perturbations
in the early universe and present-day virialized dark matter
haloes by assuming simple spherical collapse combined with
 conservation of mass and energy. We find that $z = 0$ halo concentration, 
or alternatively the inner slope of the halo density profile $\alpha$,
is monotonically and positively 
correlated with the collapse redshift of the halo. This is qualitatively similar to 
the findings of some previous works based on numerical simulations,
with which we compare our results. We then describe how the halo mass and 
concentration (or inner slope $\alpha$) can be used as two halo parameters in combination with two parameters
of early-type galaxies to create an improved abundance matching
scheme.
\end{abstract}


\section{Introduction}

Halo abundance matching is one of several methods used to link galaxies with dark matter haloes.
It uses the simple assumption that galaxy luminosity (or stellar mass) and halo mass are monotonically
related, such that more luminous galaxies reside in more massive haloes, to match 
observed magnitude-limited samples of galaxies to dark matter halo merger trees from
dark matter-only simulations \citep{valeostriker, kravtsov2004, vale2004, guo2010}.
Despite the simplicity of its underlying assumption, abundance matching is able to 
reproduce with surprising accuracy various measures related to the observed physical
distribution of galaxies, such as the luminosity functions of different cosmic environments,
the occupation numbers of haloes \citep{valeostriker},
galaxy autocorrelation functions \citep{conroy2006, guo2010, nuza2013}, 
and galaxy-galaxy lensing \citep{hearin2013, fosalba2015}.

Being able to match large samples of galaxies and haloes without the use of complex semi-analytic
or numerical hydrodynamic modeling has allowed for a simple probe of the connection between galaxies
and their dark matter haloes. Abundance matching has been used to obtain
various statistical relationships between galaxies and their host halos, such as
the luminosity-halo mass and stellar-halo mass relations \citep{vale2004, shankar2006, conroy2009, behroozi2010, 
guo2010, wake2011, leauthaud2012, moster2013}, the relationship between galaxy optical circular
velocities and the circular velocities implied from their
host haloes \citep{dutton2010}, the halo baryonic mass function \citep{baldry2008},
and the relation between central black hole mass and halo mass \citep{shankar2006}. It has also been used to examine
the dark matter haloes hosting certain types of galaxies, such as
quasars \citep{croton2009}. Furthermore, abundance matching also allows the possibility of 
assigning observed galaxies at different redshifts to simulated haloes whose 
mass growth and merger history are known, allowing one to track the evolution of galaxies through
time. This has been used to study the fate of satellite
galaxies in clusters \citep{conroy2007}, the frequency of gas-rich versus
gas-poor mergers \citep{stewart2009}, the accuracy of observational indicators
of halo mergers such as close pair counts \citep{stewart2009b}, the evolution of the
stellar-halo mass relation and velocity dispersion-halo mass relation 
and the implications for galaxy-halo co-evolution \citep{firmani2010, chae2011},
and the growth of Brightest Cluster Galaxies \citep{laporte2013}, among many others.

However, a scheme that treats luminosity as the sole
important property of a galaxy cannot be entirely correct. 
Galaxies are described by at least two parameters, as
is apparent for moderate mass systems from the bimodal distribution of galaxy colors at fixed
luminosity (e.g., \citealt{strateva2001, blanton}). This implies that at least one other
parameter of the galaxy's halo, aside from its mass, must be relevant 
to the evolution of the galaxy.

Some recent work has explored the addition of another halo and galaxy parameter 
to abundance matching at fixed stellar and halo mass 
\citep{hearinwatson2013, hearin2013, watson2014}; the authors refer to this scheme as 
``conditional abundance matching.''
In \citet{hearinwatson2013} and \citet{ hearin2013}, 
 the colors of galaxies are matched to a proxy for the halo age. The latter is
parametrized as the earliest of three times: when the main halo progenitor mass
exceeded $10^{12} M_{\odot}$, when the halo became a subhalo, 
or when the halo transitioned from fast to slow dark matter accretion, which is 
computed directly from the halo concentration as in \citet{wechsler2002}.
In practice, the last, concentration-based, age parameter is the one used for all but the most massive
galaxies. The authors find that their method is able to match a number of observables
for galaxies separated into blue and red colors, including clustering
statistics and the galaxy-galaxy lensing signal.
\citet{watson2014} apply the same method, but matching star formation rate (SFR) 
instead of galaxy color to halo age. They also report good agreement with observations, largely
due to the correlation between SFR and galaxy color.

Even considering only massive early-type galaxies, one parameter
is inadequate for predicting all of their properties. It has been known for
some time that, while a rough one-parameter relation exists
relating all variables to the velocity dispersion (the well-known
``Faber-Jackson relation;'' \citealt{faberjackson}), a two-dimensional parametrization
called the ``Fundamental Plane'' (FP) offers a superior description \citep{dressler1987}.
An evolving FP has been detected from $z = 0$ out to $z \sim 2$ \citep{sande2014}.
Other properties of early-type galaxies, 
such as observed galaxy color, as well as modeled stellar population ages and metal
abundances, have been found to be highly correlated with the FP parameters
\citep{graves1, graves2, graves3}. This implies that the properties of early-type
galaxies may be well-described by two parameters, making them a good
sample on which to test a two-parameter matching scheme.

There are also other reasons why a two-parameter matching 
scheme may work best for massive early-type galaxies.
The disk-to-bulge ratios of spiral galaxies are likely to be dependent on their environment
(e.g., \citealt{hopkins2009}).
Also, galaxy color in star-forming galaxies will not be well-correlated with 
stellar age because a recent small burst of star
 formation can make a galaxy significantly bluer while only slightly changing 
the mean stellar population age.
 
One halo parameter of physical interest is the collapse time of the dark matter halo.
That the properties of galaxies hosted by dark matter haloes could
be affected by variations in collapse time at
fixed halo mass is an idea that has appeared in a number
of previous works (e.g., \citealt{blumenthal1984}).
In particular, since the galaxy color of early-types should be
well-correlated with the age of the stars in the galaxy (as well
as their metallicity), it is possible that the galaxy color 
is also correlated with some measure of the collapse time of the host halo,
which would determine when gas could collapse and form stars.
There are also other galaxy parameters that are likely to be correlated with the 
collapse time of the host halo at fixed mass, such as the metallicity and 
stellar mass-to-light ratio. A present-day halo property correlated with
the halo collapse time could then be matched with a present-day galaxy property as a
second set of abundance matching parameters.

In this paper, we use a simple spherical collapse model to derive
a present-day halo parameter that is a proxy for the halo collapse time.
We adopt two different fitting functions for the $z = 0$ halo density profile,
and also consider physical parameters that are independent of the function used to fit
the halo profile. We will make use of the parameters we have found as part of a matching
scheme to the Fundamental Plane in a future paper.

We first derive a proxy for halo collapse time using simple spherical collapse model
in \S \ref{collapse}. We show the results of this model in \S \ref{results}.
We compare our results with previous parametrizations of halo ``formation time''
\citep{bullock2001, wechsler2002, zhao2009} derived from dark matter simulations in \S \ref{comparison}.
Finally, we describe a scheme for halo matching of the Fundamental Plane
of elliptical galaxies (a two-parameter distribution), which is not obviously
dependent on environment, that we plan to expand on in a future paper in \S \ref{future}.

In all parts of this paper we assume a cosmology consistent with the WMAP nine-year results 
plus external CMB, BAO, and $H_{0}$ measurements (\citealt{hinshaw2013}, Table 4); 
thus we take $\OM = 0.71$, $\Omega_{m, 0} = 0.29$, and $H_{0} = 69$ km/s/Mpc.

\section{Spherical Collapse Model for Halo Collapse Time}
\label{collapse}

We would like to choose some property of dark matter haloes that is a good proxy
for the halo collapse time and can be easily measured in dark matter-only simulations. 
We approximate the collapse time of a dark matter perturbation early in the universe
as twice the turnaround time in simple spherical models of collapse.  
We use a somewhat similar method to that of \citet{rubinloeb2014}, who
give equations for calculating the virialization density $\Delta_{c}$ for arbitrary pre-collapse and
post-collapse density profiles by assuming mass and energy conservation. 
Our method is similar, except that we match the initial and final
profiles within their turnaround radii at $z = 0$. 

For the purposes
of creating a two-parameter matching scheme, we choose final
halo profiles that are described by two parameters and that are
commonly used to fit numerically simulated haloes---namely, an 
NFW profile with parameters $M_{200}$ and $c$, and 
a generalized NFW profile with varying mass $M_{200}$ and inner slope $\alpha$, with
fixed $c = 5$ (see \S \ref{finalprofile}).  
We match these final profiles such that their
 mass $M$ and energy $E$ within the shell that turns around at $z = 0$ is 
the same same as that for chosen arbitrary initial profiles (e.g., tophat or Gaussian).
While this model maps final profiles to initial profiles with two parameters
uniquely (details below), the corresponding final profile is not actually the profile that the
initial profile would evolve to, as evidenced by the fact that profiles of different shapes
can be matched to the initial profiles this way. Rather, we choose final profiles
that are used as approximations for a variety of dark matter halo profile shapes.
 
\subsection{Initial Profile}
\label{initialprofile}

Here we review the equations for the evolution of the initial profile;
the analysis is similar to that in \citet{mobook}. 

We begin with some chosen initial overdensity profile $\rho_{i}(r_{i})$ at arbitrarily
chosen initial time $t_{i}$.
We assume the profile will tend to the mean matter density of the universe at that time, $\bar{\rho}(t_{i})$,
for large radii.
As long as the density is decreasing or constant with increasing radius, there will
be no shell crossing for shells that have not yet collapsed and we can treat them separately. 
It is assumed that going far enough back in time, the initial perturbation is entirely expanding, 
and none of its shells have yet turned around. 
Also, in a $\Lambda$CDM universe, a finite amount of mass will collapse in an infinite time, because 
for overdensities lower than some value, the shells expand forever due to the $\Lambda$ term instead
of collapse.
In the Appendix to this paper, we present an exact derivation of this value for an initial
profile assumed to be on the Hubble flow, which tends toward the solution having
$r = 0$ at $t = 0$ for $t_{i} \rightarrow 0$.

The collapse of each shell enclosing mass $M(<r)$ is governed by the following equation:
\begin{equation}\frac{d^{2}r}{dt^{2}} = -\frac{GM}{r^{2}} + \OM H_{0}^{2} r. \end{equation}
The above equation integrated once becomes
\begin{equation}\label{e1}\frac{1}{2}\left(\frac{dr}{dt}\right)^{2}-\frac{GM}{r}-\frac{\OM H_{0}^{2}}{2}r^{2} = \mathcal{E},\end{equation}
where $\mathcal{E}$ is the specific energy of the shell. At the turnaround time $t_{ta}$ of a given shell, this becomes 
\begin{equation}\label{e2}\mathcal{E} = -\frac{GM}{r_{ta}} - \frac{\OM H_{0}^{2}}{2}r_{ta}^{2},\end{equation}
where the turnaround radius $r_{ta}$ is the maximum radius attained by each shell.
Defining
\begin{equation}
\label{e2.5}
\zeta \equiv \frac{\OM H_{0}^{2} r_{ta}^{3}}{2GM}, \end{equation}
we see that for $\ddot{r} < 0$ at $r_{ta}$ we require $\zeta < 1/2$.
From this and Eqn. \ref{e1} we have a formula relating the radius of the shell and the time,
as long as the shell has not yet turned around:
\begin{equation}
\label{e3}
H_{0}t = \left(\frac{\zeta}{\OM}\right)^{1/2}\int_{0}^{r/r_{ta}}dx\left[\frac{1}{x}-1+\zeta(x^{2}-1)\right]^{-1/2}.\end{equation}

Using Eqn. \ref{e3}, the central collapse time ($2\times t_{ta}$) can be calculated
for $r_{ta} \rightarrow 0$ (as long as the density profile does not have a central cusp),
as can the time at which any fraction of the mass at $z=0$ collapsed.

For a selected initial density profile $\rho(r_{i})$ at time $t_{i}$,
we can find the energy of each shell from Eqn. \ref{e2}. We first 
obtain $r_{ta}$ for each $r_{i}$. To do that we re-express $\zeta$ as
\begin{equation}
\label{e4}
\zeta = \frac{\OM H_{0}^{2}}{\frac{8}{3}\pi G \rho_{i}(< r_{i})}\left(\frac{r_{ta}}{r_{i}}\right)^{3}
\end{equation}
and insert into  Eqn. \ref{e3} for $t_{i}(r_{i})$. This can be solved
numerically for
$r_{ta}$ as a function of $t_{i}$ and $r_{i}$.

By setting $t$ in Equation \ref{e3} to the age of the universe at $z = 0$
(referred to here as $t_{0}$) we can obtain $\zeta$ of the shell that is turning around at $z = 0$. 
Then we can use this same equation
to find the ratio of the original radius to the turnaround radius 
$r_{i}/r_{ta}$ of this shell by setting $t = t_{i}$. This combined with 
Equation \ref{e4} gives us the $r_{i}$ and $r_{ta}$ for the shell turning
around at $z = 0$; we will designate these as $r_{i, max}$ and $r_{ta, max}$.
We designate the mass and energy within $r_{i, max}$ at $t_{i}$ (and 
$r_{ta, max}$ at $t_{0}$) as $M_{tot}$ and $E_{tot}$, 
and these are what we will match to the mass and energy of the final
profile. We also define the initial density within $r_{i, max}$ to be $\rho_{i, max}$,
such that
\begin{equation}
\label{rmax}
r_{i, max} = \left(\frac{M_{tot}}{4/3 \pi \rho_{i, max}}\right)^{1/3}.
\end{equation}

We can find $\mathcal{E}$ for each shell using Eqn. \ref{e2}. The total energy of all the shells
within the maximum radius will then be
\begin{equation}
E_{tot} = \int_{0}^{r_{ta, max}}\mathcal{E}(r_{ta})dM(r_{ta}).
\end{equation}
To obtain $\mathcal{E}(r_{ta})$ for each shell at initial radius $r_{i}$ between $r_{i} = 0$ and $r_{i} = r_{i, max}$,
one can again insert Eqn. \ref{e4} into Eqn. \ref{e3} for $t = t_{i}$, and
solve for $r_{ta}/r_{i}$.

Scaling the size of the profile by the radius $r_{i, max}$, so that the coordinate used is
$y \equiv r/r_{i, max}$, Eqn. \ref{e2} gives
\begin{multline}
\mathcal{E} = -\frac{4}{3} \pi G \rho_{i}(<y) \frac{r_{i}}{r_{ta}} y^{2} r_{i, max}^{2} \\
-\frac{\OM H_{0}^{2}}{2}\left(\frac{r_{ta}}{r_{i}}\right)^{2} y^{2} r_{i, max}^{2}, 
\end{multline}
where $r_{ta}/r_{i}$ is a function of $y$  and also
depends on $z_{0}$ and $z_{i}$. 

Then, we can use Eqn. \ref{e4} to substitute for $r_{i}/r_{ta}$ in the equation
for $\mathcal{E}$; we obtain:
\begin{multline}
\mathcal{E} = -\frac{1}{2} \left(\frac{8}{3} \pi G \right)^{2/3} \!\! \OM^{1/3} H_{0}^{2/3} \rho_{i}^{2/3}(<y) \frac{1 + \zeta(y)}{\zeta^{1/3}(y)} y^{2} r_{i, max}^{2}.
\end{multline}
Then the total energy is then given by 
\begin{flalign}
\label{energyeq}
E_{tot} & = \int \mathcal{E} dM 
 = 4 \pi r_{i, max}^{3} \int_{0}^{1} \mathcal{E}(y) \rho_{i}(y) y^{2} dy \notag \\
& = 4 \pi r_{i, max}^{5} \left[-\frac{1}{2} \left(\frac{8}{3} \pi G \right)^{2/3} \!\! \OM^{1/3} H_{0}^{2/3} \times \right. \notag \\
& \left. \int_{0}^{1} \rho_{i}^{2/3}(<y) \rho_{i}(y) \frac{1 + \zeta(y)}{\zeta^{1/3}(y)} y^{4} dy\right],
\end{flalign}
where using Eqn. \ref{rmax} we then obtain
\begin{multline}
\label{emeq}
E_{tot}/M_{tot}^{5/3} = -\frac{3}{\rho_{i, max}^{5/3}} \frac{1}{2^{1/3}} (H_{0} G)^{2/3} \OM^{1/3} \times \\
 \int_{0}^{1} \rho_{i}^{2/3}(<y) \rho_{i}(y) \frac{1 + \zeta(y)}{\zeta^{1/3}(y)} y^{4} dy.
\end{multline}
This equation holds for any initial density profile $\rho_{i}(r_{i})$ at any chosen $z_{i}$, as long
as no part of the profile has yet collapsed.

For the initial profile, we also want to consider the limit as $z_{i}$ becomes large. Here $\delta(<y)$,
where $\rho_{i}(<y) = \bar{\rho}(z_{i})(1 + \delta(<y))$, approaches \citep{mobook}:
\begin{equation}
\label{delta}
\delta(<y) = \frac{3}{5}\frac{1 + \zeta(y)}{\zeta^{1/3}(y)}\left(\frac{\OM}{\Omega_{m, 0}}\right)^{1/3}(1+z_{i})^{-1},
\end{equation}
where $\zeta$ corresponds to a given turnaround (or collapse) time. 
We want to consider the same overdensity
profile shape at all times; i.e., $\delta(<y)/\delta_{max}$ does
not vary with time,  where $\delta_{max}$ is the overdensity at $r_{i, max}$.
It is clear that for a shell with a given collapse time, $\delta$ for that shell evolves with time;
however, the above equation shows that for large $z_{i}$, a $\delta(y)/\delta_{max}$
profile taken to be constant with time is in fact also a constant profile
in $\zeta(y)$---that is, collapse time at any fixed interior mass.

Here $\rho_{i}(y)$ changes with $z_{i}$, but for large $z_{i}$,
$\rho_{i}(y) \rightarrow \bar{\rho}(z_{i})$ as the perturbation $\delta$ decreases like
$(1+z_{i})^{-1}$ for fixed $\zeta$ (Eqn. \ref{delta}), eliminating 
$\rho_{i, max}$, $\rho_{i}(<y)$, and $\rho_{i}(y)$. This gives
\begin{multline}
\label{hiz}
E_{tot}/M_{tot}^{5/3} = -\frac{3}{2^{1/3}} (H_{0} G)^{2/3} \OM^{1/3}\frac{1 + \zeta_{0}}{\zeta_{0}^{1/3}}
 \int_{0}^{1} \frac{\delta(<y)}{\delta_{max}} y^{4} dy, 
\end{multline}
where $\zeta_{0}$ is the value of $\zeta$ for the shell that turns around at $z_{0}$, in our
case taken to be $z = 0$.

Since $\zeta_{0}$ is dependent only on the reference redshift $z_{0}$ that we select
and we assume that the overdensity profile is a constant shape, this expression
approaches a constant value as $z_{i} \rightarrow \infty$. We see that $E/M^{5/3}$ is dependent only
on the shape of the overdensity profile, which also determines the collapse times 
relative to the chosen reference redshift at high $z_{i}$. The lack of dependence
on $z_{i}$ for high $z_{i}$ is due to the fact that $\Lambda$ becomes unimportant
at early times.
 
It is also clear that for a constant or decreasing overdensity profile,
the integral on the right side of Equation \ref{hiz}
is bounded between $1/5$ and $1/2$, meaning that the energy within a shell turning
around at a given redshift must be bounded between two values.
However, as described in the next section, the final ($z = 0$) halo profiles 
we consider are common empirical fits to simulated haloes, and thus have unbounded possible
values for $E/M^{5/3}$. Therefore, not all conceivable final profiles are able 
to correspond to a possible initial profile.
 
\subsection{Final Profile}
\label{finalprofile}

We would like to find a corresponding present-day halo density profile that 
 has equal $M_{tot}$ and $E_{tot}$ within the turnaround radius at $z=0$ ($t = t_{0}$)
as the initial profile. We note that a specific turnaround
time defines a unique density within the turnaround radius at that 
time (Eqns. \ref{e2.5} and \ref{e3}), implying that for fixed $M_{tot}$,
the final and initial profiles also have the same turnaround radius. 

We choose two different forms for the present-day halo, based
on the fact that these shapes are commonly used to fit simulated haloes:
\begin{enumerate}

\item The entire profile is described by an NFW profile over the 
mean matter density:
\begin{equation}
\rho(r) = \frac{\rho_{0}}{r/r_{s}(1+r/r_{s})^{2}} + \bar{\rho}(t_{0}).
\end{equation}
As noted above, for large values of the concentration $c$, $E/M^{5/3} \propto c/(\log c)^{2}$
for the NFW profile, which is not bounded.
Therefore NFW profiles above some concentration cannot be the product of a simple spherical collapse
model.

\item While the NFW profile is the most common parametrization of dark matter halo
density, others have found that haloes are equally or better described by profiles
that have a varying inner slope (e.g, \citealt{subramanian2000}). In particular,
\citet{ricotti2007} have found that simulated dark matter haloes at virialization 
are equally well fit by either an NFW profile with varying $c$,
 or by a generalized NFW profile with fixed $c = 5$, and varying inner slope $\alpha$.
Thus we also match initial profiles to a dark matter density profile given by
\begin{equation}
\rho(r) = \frac{\rho_{0}}{(r/r_{s})^{\alpha}(1+r/r_{s})^{3-\alpha}} + \bar{\rho}(t_{0}),
\end{equation} 
where $r_{s}$ is related to $M_{200}$ via the fixed concentration $c = 5$. We refer to this
profile as an $\alpha$ model to avoid confusion with the NFW profile. This profile and
the standard NFW have the same slope of $-3$ for large $r$, making their properties
similar at large radii. As for the NFW profile, $E/M^{5/3}$ is unbounded for large $\alpha$.

\end{enumerate}

We assume haloes to be virialized within $r_{200}$, the radius within
which the mean density is 200 times the critical density, and that the virial theorem
can be used to find the energy within this region. 
Outside the virial radius, the profile is collapsing, out to a radius $r_{ta}$ at which
$t_{ta} = t_{0}$. While a different radius could be chosen within
which the profile is virialized, we note that taking the virial radius to be $r_{vir}$ as computed in 
\citet{bryannorman} would produce a negligible difference in our results.

The energy of the NFW profile within the radius turning around at $z = 0$
is the sum of the energy within the virial radius and the energy in the shells
turning around. The potential energy within the virial radius is the sum of that from the matter and that from 
$\Lambda$:
\begin{flalign}
U_{m} &= - \int_{0}^{r_{200}} 4 \pi \frac{G M(<r)}{r} \rho(r) r^{2} dr, \\
U_{\Lambda} &= - \int_{0}^{r_{200}} \frac{1}{2} \OM H_{0}^{2} 4 \pi \rho(r) r^{4} dr,
\end{flalign}
and by the virial theorem, $E(r_{200}) = 0.5U_{m} + 2U_{\Lambda}$.
Because $\bar{\rho}$ is much lower than the virial density, the fact that we
take the density profile to be an NFW or an $\alpha$ model in overdensity has an insignificant effect on
the virial energy we find within the virial radius. Thus the energies we obtain 
within the virial radius for the NFW profile
are approximately those given by equations 3.33 and 4.20 
in \citet{rubinloeb2014}.

Outside the virial radius, the region between $r_{200}$ and the 
turnaround radius $r_{ta}$ is collapsing. To determine the energy for
this region we use Eqn. \ref{energyeq}.
However, because the region is collapsing and not expanding, one must
add the turnaround time to the time a shell has been collapsing after $t_{ta}$,
thus substituting Eqn. \ref{e3} with:
\begin{multline}
H_{0}t_{0} = \left(\frac{\zeta}{\OM}\right)^{1/2}\left[\int_{0}^{1}dx\left[\frac{1}{x}-1+\zeta(x^{2}-1)\right]^{-1/2} \right. \\
\left. + \int_{r/r_{ta}}^{1}dx\left[\frac{1}{x}-1+\zeta(x^{2}-1)\right]^{-1/2}\right].
\end{multline}

For a selected initial profile shape (e.g., a Gaussian) at a chosen initial time $t_{i}$, 
the above steps will create a one-to-one mapping between initial profiles with mass $M_{tot}$
and energy $E_{tot}$ and final profiles at $z = 0$ with the same mass and energy within
the turnaround radius. Since both the 
NFW profiles and $\alpha$ model profiles
are a two-parameter family, a unique combination
of mass and energy values will correspond to a unique final profile of a given form.

\section{Results}
\label{results}

\begin{figure}
  \begin{center}
    \includegraphics[width=\columnwidth]{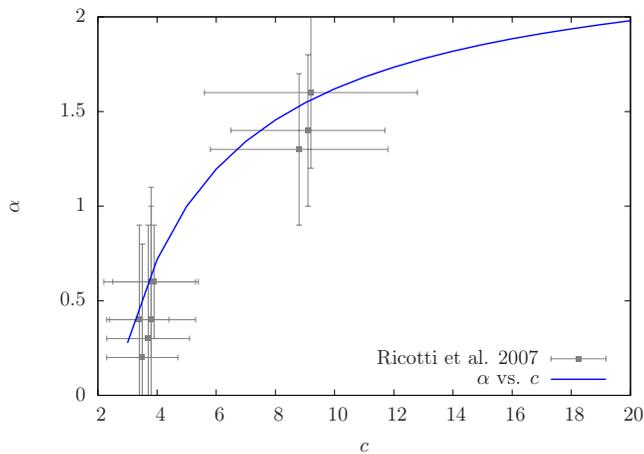} 
\caption{Correspondence between concentration $c$ of NFW profiles and inner slope of 
$\alpha$ models, for profiles at $z=0$ that have the same mass and energy within the turnaround radius. 
For comparison, gray
points are from fits to simulated haloes from \citet{ricotti2007}, described in more detail
in the text.}
    \label{fig1}
  \end{center}
\end{figure}

\begin{figure}
  \begin{center}
    \includegraphics[width=\columnwidth]{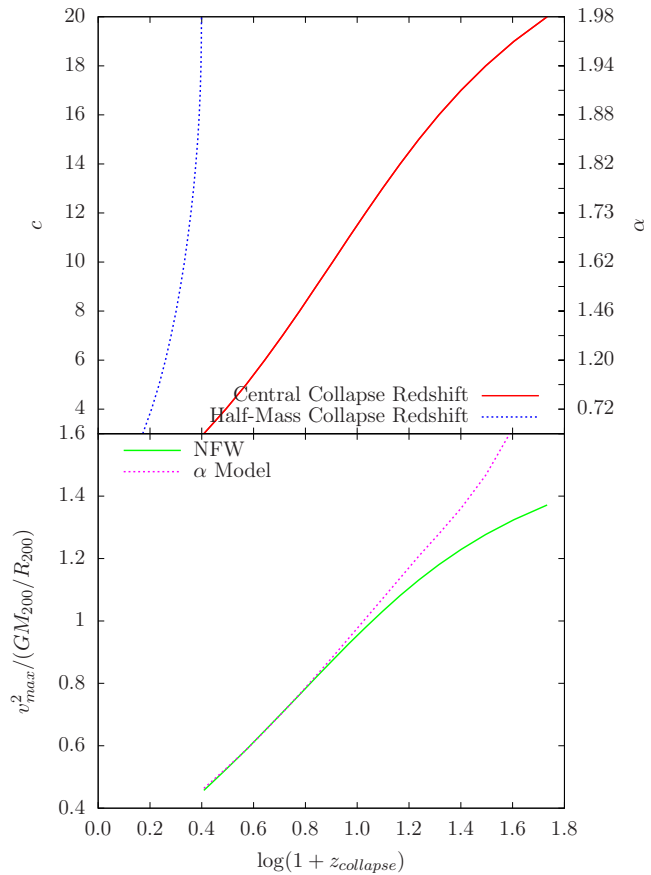} 
\caption[Figure 2]{Top Panel: The concentration $c$ of final NFW profiles, or inner slope $\alpha$ of 
$\alpha$ models, for different 
central and half-mass collapse redshifts of initial Gaussian overdensity profiles at $z_{i} = 1000$. 
The matching between initial and final profiles is described in the text. The $z = 0$ halo
concentration (or alternatively, inner slope) is monotonically and positively correlated
with the collapse redshift of any fraction of the $z = 0$ mass.

Bottom Panel: The normalized squared maximum circular velocity, 
$v^{2}_{max}/(G M_{200}/R_{200})$, at $z = 0$ for an NFW profile with varying $c$, and an $\alpha$ model with 
varying inner slope $\alpha$, versus the central collapse redshift of the matching initial Gaussian overdensity profile
with $z_{i} = 1000$. The range in $c$ and $\alpha$ shown is the same as in the top panel. 
The similarity in $v^{2}_{max}/(G M_{200}/R_{200})$ for NFW profiles and $\alpha$ models over 
the range of physical interest is apparent. }
    \label{fig2}
  \end{center}
\end{figure}

Applying the above procedure, we are able to match final profiles to initial profiles.
It is clear from Eqn. \ref{emeq} that the value $E/M^{5/3}$ at a fixed redshift is dependent only on the
shape of the overdensity profile (and the mean cosmic density at the chosen redshift), 
as it must be from dimensional arguments.
For the final profiles we take to be at $z = 0$, NFW profiles
and $\alpha$ models with the same mass and energy, this equation
implies that there will be a relationship between $c$ and $\alpha$ that
is not mass-dependent. We show this relationship in Figure 
\ref{fig1}. More concentrated
NFW profiles correspond to $\alpha$ model profiles with steeper inner slope. 
For comparison, we also show mean fits to the 40 most massive haloes
from simulations at different redshifts from \citet{ricotti2007}.
Our mapping using the energy and mass corresponds well to the match
between $\alpha$ and $c$ from direct fitting. 
This is as could be expected, since direct fitting ensures that
the profiles will have similar shapes in the region $r \sim r_{s}$,
and both profiles have slope -3 at large radii, leading to similar 
energies at fixed mass for both profile shapes.

Similarly, Equation \ref{emeq} implies that for the initial profiles,
$E/M^{5/3}$ is also a function of only the shape of the overdensity profile
and the mean density at the chosen initial redshift $z_{i}$. Furthermore, 
as seen in Equation \ref{hiz}, for a fixed overdensity shape, the 
value of $E/M^{5/3}$ approaches a constant value as $z_{i} \rightarrow \infty$.
However, the collapse time for a mass shell containing a certain fraction
of the total mass is also a function of only the chosen 
initial redshift $z_{i}$ and the shape of the overdensity
profile (Equations \ref{e3} and \ref{e4}), and also approaches a constant
value for fixed overdensity shape as $z_{i} \rightarrow \infty$ (Equation \ref{delta}).
Thus the value of $c$ or $\alpha$ of the final profile will be a function of  
the collapse time of any chosen fraction of the mass for an initial
profile of a fixed shape for fixed $z_{i}$. We show this correspondence in the top panel of Figure
\ref{fig2}, in which we present the NFW concentration $c$ 
(or $\alpha$ model inner slope) versus
the  collapse time of the center of an
initial Gaussian profile, as well as the time for half the mass
collapsed at $z=0$ of the initial profile to collapse. For the 
initial profile we take $z_{i} = 1000$ and compute the 
exact collapse times as described in \S \ref{initialprofile}; however, this redshift
is large enough that the results will be similar to those
calculated with the approximation of $z_{i} \rightarrow \infty$. 
We find that more concentrated haloes 
(or those with steeper $\alpha$) have earlier collapse times,
as might have been intuitively expected. Thus, 
either $\alpha$ or $c$ can be used as a proxy monotonically related to collapse epoch
that is independent of halo mass.

The results depend on the choice of initial profile. 
They also depend (weakly) on $z_{i}$, the 
initial redshift at which the profile is selected to be the shape of choice. This is because
in general the shape of the profile evolves over time, so that the same profile at a 
later or earlier time does not follow the same functional form. Thus the dependence on $z_{i}$ can
also be seen as equivalent to a dependence on the shape of the profile at any given time.
However, as seen in \S \ref{initialprofile}, as $z_{i} \rightarrow \infty$, the dependence
on $z_{i}$ disappears, and the value of $E/M^{5/3}$ depends only on the overdensity profile shape.

Additionally, we present two tables showing the same values as the top panel of Figure \ref{fig2};
Tables \ref{tab1} and \ref{tab2} show the central and half mass collapse time, respectively,
for NFW profiles with concentration $c$ and mass $M_{200}$. While the collapse times do not depend on $M_{200}$,
in bold we show the mass-concentration relation for simulated dark matter haloes from \citet{diemer2014},
in an observationally normalized CDM universe within which both typical values of 
$c$ and $M_{200}$ are functions of collapse epoch. We select the results from 
 \citet{diemer2014} that use the same cosmology as we assume throughout the paper
\citep{hinshaw2013}. The range in $c$ at fixed mass represents their
reported one-sigma scatter of 0.16 dex.
By nature, our calculation is done for an arbitrary initial profile, so we do not
assume cosmological initial conditions. Thus we must take the mass-concentration
relation from elsewhere. For the assumptions
about the initial profile we have made above, the bold values represent the expected
scatter in the central and half-mass collapse redshifts for NFW profiles of a given virial
mass. The ratio of the actual concentration of a halo to the mean concentration of haloes
of its virial mass, $c/\bar{c}(M_{200})$, could be used as a parameter in a two-parameter
matching scheme between haloes and galaxies that would be correlated with the halo
collapse time and potentially with galaxy properties at fixed mass, as described in our discussion of future work
in \S \ref{future}.

The bottom panel of Figure \ref{fig2} shows another parameter of the final profile,
the normalized squared maximum circular velocity $v_{max}^{2}/(GM_{200}/R_{200})$, versus the central
collapse redshift of the same initial profile as in the top panel. Again, as can be
expected from dimensional arguments, this parameter is a function of the collapse redshift.
We compare the circular velocity of the final NFW and corresponding $\alpha$ model profiles. The values are similar at 
low circular velocity (low $c$ or $\alpha$) and diverge for high circular velocity, but 
are close to one another for the relevant range of circular velocity values seen in dark matter haloes. Thus
 $v_{max}^{2}/(GM_{200}/R_{200})$ could also potentially be used as a parameter in a halo matching
scheme, and would have the benefit of not being highly dependent on the fitting function chosen for the
dark matter halo profile.

\begin{table*}
\scriptsize
\caption{Central collapse redshifts for an initial Gaussian profile at $z_{i} = 1000$.
This table shows the central collapse redshifts of initial Gaussians
matched to NFW profiles with varying $M_{200}$ and $c$. Bold numbers indicate the one-sigma
scatter around the $z = 0$ mass-concentration relation for NFWs from \citet{diemer2014}.}

\begin{tabular*}{\textwidth}{l| c c c c c c c c c}
\multicolumn{10}{c}{} \\
\toprule
\multicolumn{1}{l}{} & \multicolumn{9}{c}{$M_{200}$ [$M_{\odot}$]} \\
$c$ & 1.00e+10 & 2.51e+10 & 6.31e+10 & 1.58e+11 & 3.98e+11 & 1.00e+12 & 2.51e+12 & 6.31e+12 & 1.58e+13 \\

\midrule
3.0 & 1.56e+00 & 1.56e+00 & 1.56e+00 & 1.56e+00 & 1.56e+00 & 1.56e+00 & 1.56e+00 & 1.56e+00 & 1.56e+00 \\
4.0 & 2.11e+00 & 2.11e+00 & 2.11e+00 & 2.11e+00 & 2.11e+00 & 2.11e+00 & 2.11e+00 & 2.11e+00 & \textit{\textbf{ 2.11e+00 }} \\
5.0 & 2.73e+00 & 2.73e+00 & 2.73e+00 & 2.73e+00 & 2.73e+00 & 2.73e+00 & \textit{\textbf{ 2.73e+00 }} & \textit{\textbf{ 2.73e+00 }} & \textit{\textbf{ 2.73e+00 }} \\
6.0 & 3.40e+00 & 3.40e+00 & 3.40e+00 & 3.40e+00 & \textit{\textbf{ 3.40e+00 }} & \textit{\textbf{ 3.40e+00 }} & \textit{\textbf{ 3.40e+00 }} & \textit{\textbf{ 3.40e+00 }} & \textit{\textbf{ 3.40e+00 }}  \\
7.0 & 4.15e+00 & 4.15e+00 & 4.15e+00 & \textit{\textbf{ 4.15e+00 }} & \textit{\textbf{ 4.15e+00 }} & \textit{\textbf{ 4.15e+00 }} & \textit{\textbf{ 4.15e+00 }} & \textit{\textbf{ 4.15e+00 }} & \textit{\textbf{ 4.15e+00 }} \\
8.0 & 5.00e+00 & 5.00e+00 & \textit{\textbf{ 5.00e+00 }} & \textit{\textbf{ 5.00e+00 }} & \textit{\textbf{ 5.00e+00 }} & \textit{\textbf{ 5.00e+00 }} & \textit{\textbf{ 5.00e+00 }} & \textit{\textbf{ 5.00e+00 }} & 5.00e+00 \\
9.0 & 5.94e+00 & \textit{\textbf{ 5.94e+00 }} & \textit{\textbf{ 5.94e+00 }} & \textit{\textbf{ 5.94e+00 }} & \textit{\textbf{ 5.94e+00 }} & \textit{\textbf{ 5.94e+00 }} & \textit{\textbf{ 5.94e+00 }} & 5.94e+00 & 5.94e+00 \\
10.0 & \textit{\textbf{ 7.04e+00 }} & \textit{\textbf{ 7.04e+00 }} & \textit{\textbf{ 7.04e+00 }} & \textit{\textbf{ 7.04e+00 }} & \textit{\textbf{ 7.04e+00 }} & \textit{\textbf{ 7.04e+00 }} & 7.04e+00 & 7.04e+00 & 7.04e+00 \\
11.0 & \textit{\textbf{ 8.29e+00 }} & \textit{\textbf{ 8.29e+00 }} & \textit{\textbf{ 8.29e+00 }} & \textit{\textbf{ 8.29e+00 }} & \textit{\textbf{ 8.29e+00 }} & \textit{\textbf{ 8.29e+00 }} & 8.29e+00 & 8.29e+00 & 8.29e+00 \\
12.0 & \textit{\textbf{ 9.76e+00 }} & \textit{\textbf{ 9.76e+00 }} & \textit{\textbf{ 9.76e+00 }} & \textit{\textbf{ 9.76e+00 }} & \textit{\textbf{ 9.76e+00 }} & 9.76e+00 & 9.76e+00 & 9.76e+00 & 9.76e+00 \\
13.0 & \textit{\textbf{ 1.15e+01 }} & \textit{\textbf{ 1.15e+01 }} & \textit{\textbf{ 1.15e+01 }} & \textit{\textbf{ 1.15e+01 }} & 1.15e+01 & 1.15e+01 & 1.15e+01 & 1.15e+01 & 1.15e+01 \\
14.0 & \textit{\textbf{ 1.36e+01 }} & \textit{\textbf{ 1.36e+01 }} & \textit{\textbf{ 1.36e+01 }} & 1.36e+01 & 1.36e+01 & 1.36e+01 & 1.36e+01 & 1.36e+01 & 1.36e+01 \\
15.0 & \textit{\textbf{ 1.63e+01 }} & \textit{\textbf{ 1.63e+01 }} & \textit{\textbf{ 1.63e+01 }} & 1.63e+01 & 1.63e+01 & 1.63e+01 & 1.63e+01 & 1.63e+01 & 1.63e+01 \\
16.0 & \textit{\textbf{ 1.97e+01 }} & \textit{\textbf{ 1.97e+01 }} & 1.97e+01 & 1.97e+01 & 1.97e+01 & 1.97e+01 & 1.97e+01 & 1.97e+01 & 1.97e+01 \\
17.0 & \textit{\textbf{ 2.42e+01 }} & \textit{\textbf{ 2.42e+01 }} & 2.42e+01 & 2.42e+01 & 2.42e+01 & 2.42e+01 & 2.42e+01 & 2.42e+01 & 2.42e+01 \\
18.0 & \textit{\textbf{ 3.03e+01 }} & 3.03e+01 & 3.03e+01 & 3.03e+01 & 3.03e+01 & 3.03e+01 & 3.03e+01 & 3.03e+01 & 3.03e+01 \\
19.0 & \textit{\textbf{ 3.93e+01 }} & 3.93e+01 & 3.93e+01 & 3.93e+01 & 3.93e+01 & 3.93e+01 & 3.93e+01 & 3.93e+01 & 3.93e+01 \\
20.0 & 5.31e+01 & 5.31e+01 & 5.31e+01 & 5.31e+01 & 5.31e+01 & 5.31e+01 & 5.31e+01 & 5.31e+01 & 5.31e+01 \\
\bottomrule
\end{tabular*}
\label{tab1}
\end{table*}

\begin{table*}
\scriptsize
\caption{Half-mass collapse redshifts for an initial Gaussian profile at $z_{i} = 1000$. Same as Table \ref{tab1}, but for the half-mass collapse redshift.}
\begin{tabular}{l| c c c c c c c c c c }
\multicolumn{10}{c}{} \\
\toprule
\multicolumn{1}{l}{} & \multicolumn{9}{c}{$M_{200}$ [$M_{\odot}$]} \\
$c$ & 1.00e+10 & 2.51e+10 & 6.31e+10 & 1.58e+11 & 3.98e+11 & 1.00e+12 & 2.51e+12 & 6.31e+12 & 1.58e+13 \\

\midrule
3.0 & 4.87e-01 & 4.87e-01 & 4.87e-01 & 4.87e-01 & 4.87e-01 & 4.87e-01 & 4.87e-01 & 4.87e-01 & 4.87e-01 \\
4.0 & 6.07e-01 & 6.07e-01 & 6.07e-01 & 6.07e-01 & 6.07e-01 & 6.07e-01 & 6.07e-01 & 6.07e-01 & \textit{\textbf{ 6.07e-01 }} \\
5.0 & 7.18e-01 & 7.18e-01 & 7.18e-01 & 7.18e-01 & 7.18e-01 & 7.18e-01 & \textit{\textbf{ 7.18e-01 }} & \textit{\textbf{ 7.18e-01 }} & \textit{\textbf{ 7.18e-01 }} \\
6.0 & 8.19e-01 & 8.19e-01 & 8.19e-01 & 8.19e-01 & \textit{\textbf{ 8.19e-01 }} & \textit{\textbf{ 8.19e-01 }} & \textit{\textbf{ 8.19e-01 }} & \textit{\textbf{ 8.19e-01 }} & \textit{\textbf{ 8.19e-01 }} \\
7.0 & 9.12e-01 & 9.12e-01 & 9.12e-01 & \textit{\textbf{ 9.12e-01 }} & \textit{\textbf{ 9.12e-01 }} & \textit{\textbf{ 9.12e-01 }} & \textit{\textbf{ 9.12e-01 }} & \textit{\textbf{ 9.12e-01 }} & \textit{\textbf{ 9.12e-01 }} \\
8.0 & 9.99e-01 & 9.99e-01 & \textit{\textbf{ 9.99e-01 }} & \textit{\textbf{ 9.99e-01 }} & \textit{\textbf{ 9.99e-01 }} & \textit{\textbf{ 9.99e-01 }} & \textit{\textbf{ 9.99e-01 }} & \textit{\textbf{ 9.99e-01 }} & 9.99e-01 \\
9.0 & 1.08e+00 & \textit{\textbf{ 1.08e+00 }} & \textit{\textbf{ 1.08e+00 }} & \textit{\textbf{ 1.08e+00 }} & \textit{\textbf{ 1.08e+00 }} & \textit{\textbf{ 1.08e+00 }} & \textit{\textbf{ 1.08e+00 }} & 1.08e+00 & 1.08e+00 \\
10.0 & \textit{\textbf{ 1.15e+00 }} & \textit{\textbf{ 1.15e+00 }} & \textit{\textbf{ 1.15e+00 }} & \textit{\textbf{ 1.15e+00 }} & \textit{\textbf{ 1.15e+00 }} & \textit{\textbf{ 1.15e+00 }} & 1.15e+00 & 1.15e+00 & 1.15e+00 \\
11.0 & \textit{\textbf{ 1.22e+00 }} & \textit{\textbf{ 1.22e+00 }} & \textit{\textbf{ 1.22e+00 }} & \textit{\textbf{ 1.22e+00 }} & \textit{\textbf{ 1.22e+00 }} & \textit{\textbf{ 1.22e+00 }} & 1.22e+00 & 1.22e+00 & 1.22e+00 \\
12.0 & \textit{\textbf{ 1.28e+00 }} & \textit{\textbf{ 1.28e+00 }} & \textit{\textbf{ 1.28e+00 }} & \textit{\textbf{ 1.28e+00 }} & \textit{\textbf{ 1.28e+00 }} & 1.28e+00 & 1.28e+00 & 1.28e+00 & 1.28e+00 \\
13.0 & \textit{\textbf{ 1.34e+00 }} & \textit{\textbf{ 1.34e+00 }} & \textit{\textbf{ 1.34e+00 }} & \textit{\textbf{ 1.34e+00 }} & 1.34e+00 & 1.34e+00 & 1.34e+00 & 1.34e+00 & 1.34e+00 \\
14.0 & \textit{\textbf{ 1.38e+00 }} & \textit{\textbf{ 1.38e+00 }} & \textit{\textbf{ 1.38e+00 }} & 1.38e+00 & 1.38e+00 & 1.38e+00 & 1.38e+00 & 1.38e+00 & 1.38e+00 \\
15.0 & \textit{\textbf{ 1.42e+00 }} & \textit{\textbf{ 1.42e+00 }} & \textit{\textbf{ 1.42e+00 }} & 1.42e+00 & 1.42e+00 & 1.42e+00 & 1.42e+00 & 1.42e+00 & 1.42e+00 \\
16.0 & \textit{\textbf{ 1.46e+00 }} & \textit{\textbf{ 1.46e+00 }} & 1.46e+00 & 1.46e+00 & 1.46e+00 & 1.46e+00 & 1.46e+00 & 1.46e+00 & 1.46e+00 \\
17.0 & \textit{\textbf{ 1.48e+00 }} & \textit{\textbf{ 1.48e+00 }} & 1.48e+00 & 1.48e+00 & 1.48e+00 & 1.48e+00 & 1.48e+00 & 1.48e+00 & 1.48e+00 \\
18.0 & \textit{\textbf{ 1.49e+00 }} & 1.49e+00 & 1.49e+00 & 1.49e+00 & 1.49e+00 & 1.49e+00 & 1.49e+00 & 1.49e+00 & 1.49e+00 \\
19.0 & \textit{\textbf{ 1.50e+00 }} & 1.50e+00 & 1.50e+00 & 1.50e+00 & 1.50e+00 & 1.50e+00 & 1.50e+00 & 1.50e+00 & 1.50e+00 \\
20.0 & 1.50e+00 & 1.50e+00 & 1.50e+00 & 1.50e+00 & 1.50e+00 & 1.50e+00 & 1.50e+00 & 1.50e+00 & 1.50e+00 \\
\bottomrule
\end{tabular}
\label{tab2}
\end{table*}

\begin{figure}
  \begin{center}
    \includegraphics[width=\columnwidth]{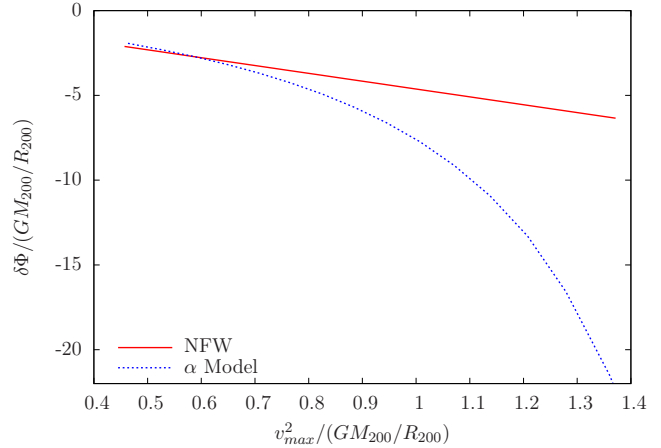} 
\caption{The normalized squared maximum circular velocity, $v^{2}_{max}/(G M_{200}/R_{200})$,
 versus the normalized excess central potential, $\delta \Phi/(G M_{200}/R_{200})$,
at $z = 0$ for an NFW profile with varying $c$ and an $\alpha$ model with 
varying inner slope $\alpha$. Unlike the maximum circular velocity shown in Figure \ref{fig2},
the central potential is significantly different for the two models.}
    \label{fig5}
  \end{center}
\end{figure}

In Figure \ref{fig5}, we show the $v_{max}^{2}/(G M_{200}/R_{200})$ versus a related parameter,
the excess normalized central potential $\delta \Phi/(G M_{200}/R_{200})$, for both NFW profiles
and $\alpha$ models. While the values of $v_{max}^{2}/(G M_{200}/R_{200})$ are similar for both
models at fixed mass and energy, the values of the central potential are significantly different
for the two profiles shapes, implying that the maximum circular velocity is a superior 
``common'' parameter between the two types of models to use to predict the initial halo
collapse redshift. 

\begin{figure}
  \begin{center}
    \includegraphics[width=\columnwidth]{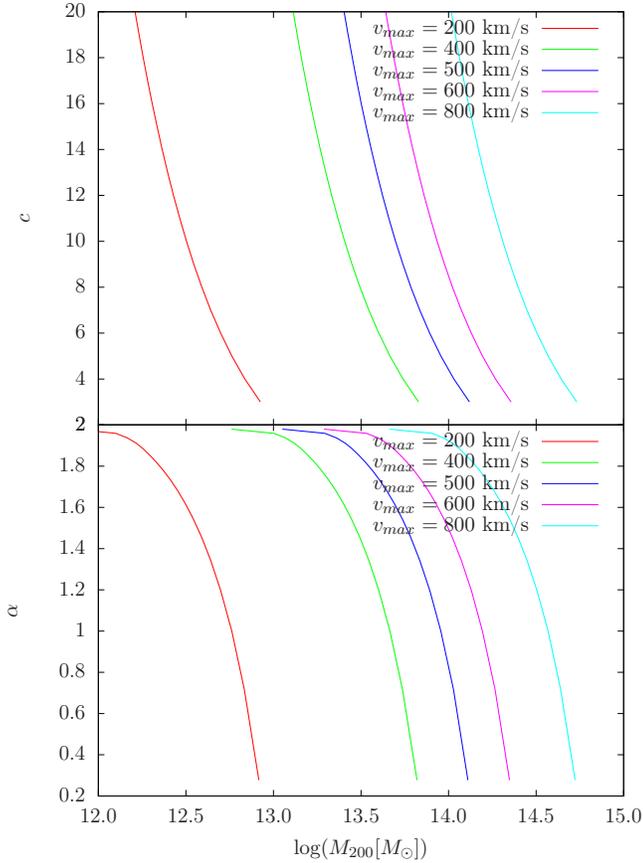} 
\caption{Contours of constant $v_{max}$ for varying $M_{200}$ and $c$ for NFW profiles in the top panel, and the 
same for varying $M_{200}$ and inner slope for $\alpha$ models in the bottom panel.}
    \label{fig6}
  \end{center}
\end{figure}

For reference, 
we show contours of constant $v_{max}$ in Figure \ref{fig6},
in the top panel for varying $M_{200}$ and $c$ for NFW profiles, and in the bottom panel
for varying $M_{200}$ and $\alpha$ for $\alpha$ models, over the range of physical interest.

\section{Comparison to Previous Work}
\label{comparison}

Previous papers have investigated parametrizations of the time at which a halo formed,
usually based on the results of dark matter simulations.  Such papers
include \citet{bullock2001} and \citet{wechsler2002}. The model of \citet{wechsler2002} improved upon
that of \citet{bullock2001}, but obtained similar results.

\citet{wechsler2002} examined the mass accretion histories of the most massive 
progenitors of individual haloes from dark matter simulations for $z < 7$, defining the halo mass to be the mass
within $\Delta_{\mathrm{vir}}$ \citep{bryannorman}.
\citet{wechsler2002} found that the mass accretion histories of haloes at each timestep, 
despite halo mergers in the simulation, can generally be fit well by a simple analytic form:
\begin{equation}
\label{eq1}
M(a) = M_{0}\exp\left[-a_{c} S \left( \frac{a_{0}}{a} - 1 \right)\right].
\end{equation}
Here $a$ is the scale factor, $a_{0}$ is a reference scale factor at which the mass of
the halo is $M_{0}$, $S$ is an arbitrarily chosen constant, and $a_{c}$ is taken
to be the ``formation scale factor'' of the halo, given the choice of $S$. This form
 is self-consistent for different choices of $a_{0}$.
Taking the formation scale factor $a_{c}$ of the halo as defined above with $S$ chosen to be 2, 
the authors assign a formation
time to each halo and find that the concentration of each halo at some reference scale 
factor $a_{0}$ is related to the
formation time via
\begin{equation}
\label{eq1.1}
c_{\mathrm{vir}} = 4.1 a_{0}/a_{c}
\end{equation}
where the constant 4.1 is the concentration of haloes ``forming'' at the present day, given the
choice of $S = 2$.

A more recent paper looking at mass accretion histories in dark matter simulations
 is \citet{zhao2009}. The authors claim to obtain more accurate fits to halo accretion
histories using a more complex model than \citet{wechsler2002} and \citet{bullock2001}.
Unlike \citet{wechsler2002}, they find that mass accretion histories follow a power-law
form as opposed to an exponential.
\citet{zhao2009} find that the concentration of haloes at some 
observation time $t$ is a function of the time at which their mass reached $4\%$ of the
mass at the observation time, given by
\begin{equation}
c = [4^{8} + (t/t_{0.04})^{8.4}]^{1/8}
\end{equation}
where $t_{0.04}$ is the time at which $4\%$ of the mass at $t$ was reached.
Note that halo concentrations in this model cannot be less than 4.

Due to the fact that we begin with arbitrary initial conditions and not
cosmological ones, it is difficult to compare our results with those of \citet{wechsler2002}
and \citet{zhao2009}. However, both our model and the results obtained by these two
papers find a monotonic relation between the collapse time or ``formation time'' of 
haloes and their concentrations at fixed mass at any given time, where a higher concentration
implies an earlier formation time. This is crucial as abundance matching-type methods
require a monotonic relation between the parameters being matched.

\section{Future Work}
\label{future}

Using a combination of two halo parameters, one can perform
two-parameter abundance matching to two observable parameters of galaxies. 
As shown above, parameters that correlate well with the physically relevant
parameter of halo collapse time include $c$ or $\alpha$, or alternatively
$v_{max}^{2}/(GM_{200}/R_{200})$. These could be combined with the parameter
$M_{200}$ used in standard abundance matching, although these are not the only
possible choices for matching.

Because we only
consider two parameters, and standard abundance matching is most effective
for massive galaxies, we plan to focus on two-parameter abundance matching
to the Fundamental Plane of massive early type galaxies in future work. 
The Fundamental Plane is an observed 
relationship between three parameters: 
the effective radius $R_{e}$, the velocity dispersion
$\sigma$, and the surface brightness within $R_{e}$, $I_{e}$ (or alternatively, the luminosity
within $R_{e}$). Elliptical galaxies occupy a plane in the space of these three parameters: 
 \begin{equation}
\log R_{e} = a \log \sigma + b \log I_{e} + c_{z},
\end{equation}
where $a$ and $b$ are constants and $c_{z}$ is a redshift-dependent zero-point.
Assuming that the mass-to-light ratio of elliptical galaxies is roughly independent of stellar mass,
and that the galaxies are fully virialized, one would expect from the virial theorem
that the fundamental plane would follow $R_{e} \propto \sigma^{2}I_{e}^{-1}$. However,
the observed fundamental plane has a tilt with respect to this relation (e.g., \citealt{bezanson2013}).
Although the FP has some thickness and is not completely two-dimensional,
this thickness is small \citep{graves2, graves3}.
While we plan to focus on early-type galaxies, interestingly,
\citet{bezanson2014} find that early-
and late-type galaxies fall on the same mass fundamental plane.

Additionally, other galactic parameters have been found to be strongly correlated with 
parameters of the Fundamental Plane. These include galaxy color, stellar mass-to-light ratio, 
and mean stellar population age and metallicity \citep{graves1, graves2, graves3, porter2014}.
Thus matching to the FP would take into account the variance in these parameters as well,
potentially allowing a near-complete prediction of the properties of a massive galaxy residing
in a halo based off the halo's properties, and creating a better connection between the properties
of the halo (including formation epoch) and the properties of the galaxy.

\makeatletter
\let\jnl@style=\rmfamily 
\def\ref@jnl#1{{\jnl@style#1}}% 
\newcommand\aj{\ref@jnl{AJ}}%        % Astronomical Journal 
\newcommand\araa{\ref@jnl{ARA\&A}}%  % Annual Review of Astron and Astrophys 
\newcommand\apj{\ref@jnl{ApJ}}%    % Astrophysical Journal 
\newcommand\apjl{\ref@jnl{ApJ}}%     % Astrophysical Journal, Letters 
\newcommand\apjs{\ref@jnl{ApJS}}%    % Astrophysical Journal, Supplement 
\newcommand\ao{\ref@jnl{Appl.~Opt.}}%   % Applied Optics 
\newcommand\apss{\ref@jnl{Ap\&SS}}%  % Astrophysics and Space Science 
\newcommand\aap{\ref@jnl{A\&A}}%     % Astronomy and Astrophysics 
\newcommand\aapr{\ref@jnl{A\&A~Rev.}}%  % Astronomy and Astrophysics Reviews 
\newcommand\aaps{\ref@jnl{A\&AS}}%    % Astronomy and Astrophysics, Supplement 
\newcommand\azh{\ref@jnl{AZh}}%       % Astronomicheskii Zhurnal 
\newcommand\baas{\ref@jnl{BAAS}}%     % Bulletin of the AAS 
\newcommand\icarus{\ref@jnl{Icarus}}% % Icarus
\newcommand\jrasc{\ref@jnl{JRASC}}%   % Journal of the RAS of Canada 
\newcommand\memras{\ref@jnl{MmRAS}}%  % Memoirs of the RAS 
\newcommand\mnras{\ref@jnl{MNRAS}}%   % Monthly Notices of the RAS 
\newcommand\pra{\ref@jnl{Phys.~Rev.~A}}% % Physical Review A: General Physics 
\newcommand\prb{\ref@jnl{Phys.~Rev.~B}}% % Physical Review B: Solid State 
\newcommand\prc{\ref@jnl{Phys.~Rev.~C}}% % Physical Review C 
\newcommand\prd{\ref@jnl{Phys.~Rev.~D}}% % Physical Review D 
\newcommand\pre{\ref@jnl{Phys.~Rev.~E}}% % Physical Review E 
\newcommand\prl{\ref@jnl{Phys.~Rev.~Lett.}}% % Physical Review Letters 
\newcommand\pasp{\ref@jnl{PASP}}%     % Publications of the ASP 
\newcommand\pasj{\ref@jnl{PASJ}}%     % Publications of the ASJ 
\newcommand\qjras{\ref@jnl{QJRAS}}%   % Quarterly Journal of the RAS 
\newcommand\skytel{\ref@jnl{S\&T}}%   % Sky and Telescope 
\newcommand\solphys{\ref@jnl{Sol.~Phys.}}% % Solar Physics 
\newcommand\sovast{\ref@jnl{Soviet~Ast.}}% % Soviet Astronomy 
\newcommand\ssr{\ref@jnl{Space~Sci.~Rev.}}% % Space Science Reviews 
\newcommand\zap{\ref@jnl{ZAp}}%       % Zeitschrift fuer Astrophysik 
\newcommand\nat{\ref@jnl{Nature}}%  % Nature 
\newcommand\iaucirc{\ref@jnl{IAU~Circ.}}% % IAU Cirulars 
\newcommand\aplett{\ref@jnl{Astrophys.~Lett.}}%  % Astrophysics Letters 
\newcommand\apspr{\ref@jnl{Astrophys.~Space~Phys.~Res.}}% % Astrophysics Space Physics Research 
\newcommand\bain{\ref@jnl{Bull.~Astron.~Inst.~Netherlands}}% % Bulletin Astronomical Institute of the Netherlands 
\newcommand\fcp{\ref@jnl{Fund.~Cosmic~Phys.}}%   % Fundamental Cosmic Physics 
\newcommand\gca{\ref@jnl{Geochim.~Cosmochim.~Acta}}% % Geochimica Cosmochimica Acta 
\newcommand\grl{\ref@jnl{Geophys.~Res.~Lett.}}%  % Geophysics Research Letters 
\newcommand\jcap{\ref@jnl{JCAP}}
\newcommand\jcp{\ref@jnl{J.~Chem.~Phys.}}%     % Journal of Chemical Physics 
\newcommand\jgr{\ref@jnl{J.~Geophys.~Res.}}%     % Journal of Geophysics Research 
\newcommand\jqsrt{\ref@jnl{J.~Quant.~Spec.~Radiat.~Transf.}}%   % Journal of Quantitiative Spectroscopy and Radiative Trasfer 
\newcommand\memsai{\ref@jnl{Mem.~Soc.~Astron.~Italiana}}% % Mem. Societa Astronomica Italiana 
\newcommand\nphysa{\ref@jnl{Nucl.~Phys.~A}}%     % Nuclear Physics A 
\newcommand\physrep{\ref@jnl{Phys.~Rep.}}%       % Physics Reports 
\newcommand\physscr{\ref@jnl{Phys.~Scr}}%        % Physica Scripta 
\newcommand\planss{\ref@jnl{Planet.~Space~Sci.}}%  % Planetary Space Science 
\newcommand\procspie{\ref@jnl{Proc.~SPIE}}%      % Proceedings of the SPIE 
\let\astap=\aap 
\let\apjlett=\apjl 
\let\apjsupp=\apjs 
\let\applopt=\ao 
\makeatother

\section*{Acknowledgments}
We thank Sandra Faber, Alexie Leauthaud, and Renyue Cen for their comments on the paper. A.K.
was supported by the National Science Foundation Graduate Research Fellowship,
Grant No. DGE-1148900.

\bibliographystyle{mn2e}

\section*{Appendix}
\label{appendix}

In \S \ref{initialprofile}, we show the equations determining the evolution
of mass shells of an overdensity in a $\Lambda$CDM universe, where each shell has initial
condition $r=0$ at $t=0$. In a $\Lambda$CDM universe, mass shells
with overdensity $\delta$ below some value will never turn around. For the 
case presented in \S \ref{initialprofile}, there is no simple analytic formula
for this overdensity for all times (although it can be found numerically 
from the equations given therein), but there is one as the initial
time approaches $0$, given by Eqn. \ref{delta}. However, for the case in which all mass shells
are taken to be initially on the 
Hubble flow (i.e., $\dot{r}_{i} = H_{i}(t_{i})r_{i}$ for a chosen initial time $t_{i}$),
there is an analytic solution for the minimum overdensity to turn around for
all initial times. Here we present a short derivation of this value and
compare to the solution for initial conditions $r=0$ at $t=0$ at early times.

Once again, we begin with the equation of motion for a shell containing mass $M$:
\begin{equation}\frac{d^{2}r}{dt^{2}} = -\frac{GM}{r^{2}} + \OM H_{0}^{2} r. \end{equation}
Integrated once, we obtain
\begin{equation}
\label{ee2}
\frac{1}{2}\dot{r}^{2} - \frac{1}{2}\dot{r}_{i}^{2}
= \frac{GM}{r} - \frac{GM}{r_{i}} + \frac{\OM H_{0}^{2}}{2}r^{2} -  \frac{\OM H_{0}^{2}}{2}r_{i}^{2}.\end{equation}
Here our initial condition is 
\begin{equation}
\dot{r}_{i}^{2} = H_{i}^{2}r_{i}^{2}  = \frac{2G\langle M \rangle}{r_{i}} + \OM H_{0}^{2} r_{i}^{2}, 
\quad \langle M \rangle \equiv \frac{4}{3} \pi G \bar{\rho}(t_{i}) r_{i}^{3}.
\end{equation}
Combined with Eqn. \ref{ee2}, we obtain
\begin{equation}
\dot{r}^{2} = \frac{2GM}{r} + \OM H_{0}^{2} r^{2} - \frac{2G(M - \langle M \rangle)}{r_{i}}.
\end{equation}
Now we define
\begin{equation}
r/r_{i} \equiv \chi, \quad \frac{d}{dt} \equiv H_{i} \frac{d}{d\tau}
\end{equation}
giving
\begin{equation}
\left(\frac{d\chi}{d\tau}\right)^{2} = \frac{8\pi G \bar{\rho}}{3 H_{i}^{2}} \frac{M}{\langle M \rangle} \frac{1}{\chi}
+ \frac{\OM H_{0}^{2}}{H_{i}^{2}} \chi^{2} - \frac{8\pi G \bar{\rho}}{3 H_{i}^{2}} \frac{M - \langle M \rangle}{\langle M \rangle}.
\end{equation}
Noting that
\begin{equation}
1 + \delta = \frac{M}{\langle M \rangle}, \quad \frac{8\pi G \bar{\rho}}{3 H_{i}^{2}} = \Omega_{m, i}, \quad \frac{\OM H_{0}^{2}}{H_{i}^{2}} = \Omega_{\Lambda, i}
\end{equation}
and that for a flat universe $\Omega_{m, i} + \Omega_{\Lambda, i} = 1$,
we obtain
\begin{equation}
\left(\frac{d\chi}{d\tau}\right)^{2} = \frac{\Omega_{m, i} (1+\delta)}{\chi} + \Omega_{\Lambda, i}\chi^{2} - \Omega_{m,i}\delta.
\end{equation}
For any shell that turns around, $(d\chi/d\tau)^{2} = 0$ for some 
maximal value of $\chi$. Therefore we want to find the smallest $\delta$ such that
there exists a positive $\chi$ where
\begin{equation}
\chi^{3} - \frac{\Omega_{m, i}}{\Omega_{\Lambda, i}}\delta\chi + \frac{\Omega_{m, i}}{\Omega_{\Lambda, i}}(1 + \delta) = 0.
\end{equation}
We also require that this cubic to be positive for $\chi$ between
$0$ and the maximum value of $\chi$ in order for $d\chi/d\tau$ to be a real number.
This cubic has inflection points at $\chi_{\pm} = \pm(\Omega_{m, i} \delta /3 \Omega_{\Lambda, i})^{1/2}$;
for the previously mentioned conditions to hold, it requires that the value of the cubic at $\chi_{+}$
be negative or $0$:
\begin{equation}
\left(\frac{\Omega_{m, i}}{\Omega_{\Lambda, i}}\frac{\delta}{3}\right)^{3/2} - 
\frac{1}{3^{1/2}}\left(\frac{\Omega_{m, i}}{\Omega_{\Lambda, i}} \delta \right)^{3/2} + \frac{\Omega_{m, i}}{\Omega_{\Lambda, i}} 
(1 + \delta) \leq 0.
\end{equation}
This reduces to simply
\begin{equation}
\frac{\delta^{3}}{(1 + \delta)^{2}} \geq \frac{27}{4} \frac{\Omega_{m, i}}{\Omega_{\Lambda, i}}.
\end{equation}
As stated, this is the exact solution for an initial profile that is assumed to be on the Hubble flow
at time $t_{i}$. This can be compared to Equation \ref{delta}, which is the solution for $\delta$ 
with initial conditions $r = 0$ at $t = 0$ for all shells as $z_{i} \rightarrow \infty$. Here
the boundary between collapsing and forever expanding shells is given by $\zeta = 0.5$, by the definition
of $\zeta$. This gives for the minimal overdensity at early times
\begin{equation}
\delta = \frac{9}{10}2^{1/3} \left(\frac{\Omega_{m, i}}{\Omega_{\Lambda, i}} \right)^{1/3};
\end{equation}
whereas at early times for the solution on the Hubble flow we find
\begin{equation}
\delta \approx \frac{3}{2}2^{1/3}\left(\frac{\Omega_{m, i}}{\Omega_{\Lambda, i}} \right)^{1/3}.
\end{equation}
It can be seen that the derived $\delta$ differs by a value of $3/5$. This is the difference in
overdensity expected for a matter-dominated universe between considering a hypersurface of constant
$H_{i}$ and one of constant $t_{i}$, as derived in \citet{gottgunn}. This ratio between the results will not
hold at small $z_{i}$ due to the effect of $\Lambda$, but for early times the universe is 
increasingly matter dominated and so the two solutions are the same modulo
this factor of $3/5$.

\label{lastpage}

\end{document}